\documentstyle[psfig,conf_iap,]{article}
\begin{document}
\heading{%
%
The Cosmic Coincidence Problem\\
in the Brane World\\

%
} 
\par\medskip\noindent
\author{%
Massimo Pietroni$^{1}$
}
\address{%
INFN, sezione di Padova,
via F.~Marzolo 8, I-35131 Padua, Italy.
}

\begin{abstract}
The so called  `cosmic coincidence' problem seems to suggest a coupling between dark energy and dark matter. A possibility could be given by dark matter particles with masses depending exponentially on the scalar field associated to dark energy.  If the latter also has an exponential potential, attractor solutions exist with constant $\Omega_{DM}$ and $\Omega_{DE}$ and negative effective equation of state.
Due to the very low mass of the dark energy field, its coupling to baryons should be strongly suppressed. 
Here we present a natural realization of this scenario in a model with large extra dimensions.

%
\end{abstract}
\def\beqra{\begin{eqnarray}}
\def\eeqra{\end{eqnarray}}
\def\beq{\begin{equation}}               
\def\eeq{\end{equation}}
\def\ds{\displaystyle}
\def\ts{\textstyle}
\def\ss{\scriptstyle}
\def\sss{\scriptscriptstyle}
\def\Vb{\bar{V}}
\def\phb{\bar{\phi}}
\def\rhb{\bar{\rho}}
\def\L{\Lambda}
\def\T{\Theta}
\def\re#1{(\ref{#1})}
\def\D{\Delta}
\def\G{\Gamma}
\def\p{\partial}
\def\half{\mbox{\small$\frac{1}{2}$}}  
\def\de{\delta}
\def\mm{\rm mm}
\def\mic{\mu {\rm m}}
\def\i{i}
 \def\f{f}
 \def\d{d}
 \def\e{e}
\def\lta{\mathrel{\vcenter{\hbox{$<$}\nointerlineskip\hbox{$\sim$}}}}
\def\gta{\mathrel{\vcenter{\hbox{$>$}\nointerlineskip\hbox{$\sim$}}}}
\renewcommand{\Re}{\mathop{\mathrm{Re}}}
\renewcommand{\Im}{\mathop{\mathrm{Im}}}
\newcommand{\tr}{\mathop{\mathrm{tr}}}
\newcommand{\Tr}{\mathop{\mathrm{Tr}}}

\section{Coincidence Problem {\it vs} Equation of State}
One of the most puzzling aspects of the current cosmological paradigm, is the so-called `cosmic coincidence problem', {\it i.e.} the quasi equivalence between dark matter (DM) and dark energy (DE) densities today. 
Indeed, for a non-interacting fluid $A$, the continuity equation $d(\rho_A a^3) = - p_A da^3$,
relates the equation of state $w_A \equiv p_A/\rho_A$ to the evolution of the energy density with the expansion, {\it i.e.}, $ \rho_A \sim a^{-3(w_A+1)}$.
Therefore, the ratio between dark matter ($w_{DM}=0$) and a dark-energy component with negative pressure ($w_{DE}<0$) scales as
\beq
\frac{\rho_{DM}}{\rho_{DE}} \sim a^{3 w_{DE}}\;,
\label{ratio}
\eeq 
and could be only `coincidentally' of order one today. This is the case for the cosmological constant ($w_{DE}=-1$), for a scalar field with an inverse power law potential $V(\varphi)\sim \varphi^{-n}$ ($w_{DE}=-2/(n+2)$), or with any potential allowing solutions with negative pressure. On the other hand, exponential potentials exhibit attractor solutions in which there is no cosmic coincidence problem, since the scalar field energy density scales at a constant ratio with the background, be it matter or radiation. But, by virtue of eq.~(\ref{ratio}) this implies  $w_{DE}=0$ after equivalence, thus making it unacceptable as a DE component.
Thus, there seems to be an inextricable contradiction between a negative equation of state and the elimination of the cosmic coincidence.

The impasse may be escaped by relaxing the assumption leading to eq.~(\ref{ratio}), {\it i.e.} by allowing an interaction between DM and DE, besides the gravitational one.
\section{Exponential Vamps}
An interesting possibility is to consider dark matter particles whose masses depend on the quintessence field, or VArying Mass ParticleS (Vamps), which were considered in \cite{vamp}. Assuming exponential dependencies both in the potential and in the DM particle mass,
\beq
V(\varphi)=V_0 \exp(\beta \varphi)\;,\;\;\;\;\; M=M_0 \exp{(-\lambda \varphi})\;,
\label{exp}
\eeq
an {\it attractor} solution exists on which both $\rho_{\mathrm Vamps}$ and $\rho_{DE}$ scale as $a^{-3(W+1)}$, with
\beq
W=-\frac{\lambda}{\lambda+\beta}\;,\;\;\;\;\;\;\;{\mathrm and}\;\;\;\;\Omega_{DE}=1-\Omega_{\mathrm Vamps}=\frac{3+\lambda(\lambda+\beta)}{(\beta+\lambda)^2}\;,
\label{attr}
\eeq
provided $\beta > \frac{-\lambda +\sqrt{\lambda^2+12}}{2}$, thus solving the coincidence problem with a negative effective equation of state.

It should be noted that the crucial requirement in order to get an attractor with the above properties, is to have {\it exponential} dependencies both in the potential and in the mass, a case not  considered in refs. \cite{vamp}. Moreover, the coupling $\lambda$ between Vamps and $\varphi$ should be $O(1)$ in order to have a sensibly negative equation of state. If also baryons were coupled so strongly with the almost massless $\varphi$, we would immediatly run into problem with the bounds coming from tests of the fifth force or the equivalence principle. In the remaining part of my talk, I will discuss how a model exhibiting these properties \cite{io} can be formulated in scenarios with large extra dimensions \cite{arkani}.

Before doing that, let me mention that a similar model has been discussed recently in \cite{amendola}. Those authors implement the DE-DM interaction by assuming non-vanishing divergences for the energy momentum tensors 
\beq
(T_{DM}^{\mu\nu})_{;\mu} = -(T_{DE}^{\mu\nu})_{;\mu} = \lambda T_{DM} g^{\nu\rho}\varphi,_{\rho}\;,\;\;\;\;\;({\mathrm Amendola}\;\;et\;\;\;al)
\eeq
where $T_{DM}$ is the trace of the $DM$ energy-momentum tensor. The background behavior turns out to be the same as (\ref{attr}), but fluctuations evolve differently, since in the Vamps case we have
\beq
(T_{DM}^{\mu\nu})_{;\mu} = -(T_{DE}^{\mu\nu})_{;\mu} = \lambda T_{DM} u^\nu u^\rho \varphi,_{\rho}\;\;\;({\mathrm for \;\;Vamps}),
\eeq 
where $u^\mu$ is the four-velocity of Vamps particles. This will be studied in a forthcoming paper.

\section{Vamps and Large Extra Dimensions}
In principle, brane world scenarios offer a suggestive contact with the DE
problem. Indeed, extra spatial  dimensions compactified to a size as
large as $100 \mic - 1 \mm$ have been shown to be a  viable
possibility, provided that no Standard Model (SM) field propagates through
them \cite{arkani}. 
It might then  be tempting to attribute the observed value of
the DE energy density to the Casimir energy  associated to some field
propagating in these large extra dimensions of size $r$, {\it i.e.}
\beq
\rho_{\mathrm Casimir} \simeq 1/r^4.
\label{rhode}
\eeq 

If the radius $r$ is stabilized, then the
energy density behaves  exactly as a cosmological constant, and we
have made no substantial progress with respect to the cosmic coincidence problem. 
It seems then necessary to make the radius
$r$ dynamical, so that it may evolve on a cosmological time-scale. 
But this turns out to be quite dangerous. 
The `radion' field, whose expectation value fixes $r$, is ultra-light ($m\sim H_0\simeq 10^{-33}{\mathrm eV}$) and
couples to the trace of the SM energy momentum tensor  with gravitational
strength. It then behaves
as a practically massless Brans-Dicke scalar with $O(1)$ couplings to matter,
whereas present bounds are $O(10^{-3})$ \cite{will}.

However,  besides 
large extra-dimensions in the $100 \,\mic$ range there could also be
smaller ones, and this may solve the problem. Considering for simplicity 
only two compact subspaces, each caracterized by a single radius, we note that
the trace of the 
four-dimensional
energy-momentum tensor couples with the total {\it volume} of the compact 
space, that
is with the combination $r_s^{n_s} r_l^{n_l}$, where $r_{s,l}$ and $n_{s,l}$ 
are the radii and dimensionalities of the two subvolumes, 
respectively. 
If we assume a
stabilization mechanism that fixes the {\it total volume} ${\cal V}$
 of the compactified
manifold but {\it not its shape}, the potentially dangerous combination of 
radion fields associated
with volume fluctuations is made harmless, whereas the orthogonal one, 
associated with
shape deformation, is decoupled from normal matter and may then 
be ultra-light.
As a consequence, $r_l$ can grow on a cosmological
time-scale (such that the associated Casimir energy $1/r_l^4$
decreases) and at the same time $r_s$ shrinks so that ${\cal V}$ keeps a fixed
value.

If we now identify as DM candidate some stable state living on the
$4+n_s$-dimensional brane, we may obtain a realization of the general Vamps model of eq.~(\ref{exp}). Indeed, if the mass of the DM particle is inversely proportional 
to some power of $r_s$, as for a Kaluza-Klein (KK)
state ($m_{DM} = m_{KK} \sim r_s^{-1}$), then its energy density scales as
\beq
\rho_{DM} \sim 1/(r_s a^3)  
\;.
\label{rhodm}
\eeq
Assuming that the total volume of the compact subspace is fixed, the Casimir energy
(\ref{rhode}) and the DM energy  density (\ref{rhodm}) give
\beq
V( \phi ) \sim
\exp\left(4 \sqrt{\frac{n_s}{n_l {\cal N}}} \phi\right)\,,\;\;\;{\mathrm and} \;\;
\rho_{DM} \sim
\exp\left(- \sqrt{\frac{n_l}{n_s {\cal N}}}
\phi_2\right) a^{-3}\,,
\eeq
respectively, where $\phi$ is the canonically normalized field associated to shape deformations of the compact sub-manifold \cite{io}. Moreover, ordinary matter such as baryons is only sensitive to the global volume of the compact manifold, not to its shape, and is thus decoupled from $\phi$.

In conclusion, we have realized a Vamps model in which the DM-DE coupling is naturally $O(1)$, and the DE-baryon coupling is naturally suppressed. The parameters $\lambda$ and $\beta$  of eq.~(\ref{exp}) are fixed by geometry, {\it i.e.} by $n_l$ and $n_s$, and then so are
 the values of $W$ and $\Omega_{DE}$ on the attractor.

In Tab.~1 we list the possible values
of $\Omega_{DE}$, $W$ and of 
$H_0 t_0$, $t_0$ being the present age of the Universe. 
We limited the dimensionality of
the compact space according to the theoretical prejudice coming from string 
theory, {\it i.e.} $n_l+n_s\le 6$.

\begin{center}
\begin{tabular}{||c|c|c|c|c||}

\multicolumn{5}{l}{{\bf Table 1.} Possible attractors for $n_l+n_s\le 6$} \\
\hline
$n_s$ & $n_l$ & $\Omega_{DE}$ & $W$ & $H_0 t_0$\\
\hline
\hline
1 - 5 &   1 &      $\le 0.44$  &  --  & -- \\
\hline
1     &   2 &  0.83            & $-1/3$ & 1 \\
\hline
2     &   2 &  0.68            &  -0.20& 0.83\\
\hline
3     &   2 &  0.60            &  -0.14& 0.78\\
\hline
4     &   2 &  0.56            & -0.11 & 0.75\\
\hline
1-3   &   3 &  $\ge 0.92$      &  --     &   --  \\
\hline
1-2   &   4 & no attractor& -- & -- \\
\hline
1    &5& no attractor & -- & -- \\
\hline
\hline
\end{tabular}  
\end{center}

A noticeable fact about Tab.~1 is that the observed range for
the dark energy density, $0.6 \lta \Omega_{DE} \lta 0.8$ 
uniquely selects the number of `large' extra dimensions to be $n_l=2$, the
same value that is required by the totally unrelated issue of solving the
hierarchy problem with `millimeter' size extra dimensions \cite{arkani}.

A model-independent prediction of the framework outlined
in this paper is the presence of extra 
dimensions in the $100\, \mic$ range. Whereas for the solution of the hierarchy
problem discussed in \cite{arkani} this value was {\it allowed}, the link 
with the cosmological expansion discussed in this paper makes it a true
{\it prediction}. Even more interestingly, the expected value 
for $r_l$ is quite close to the present bound of
$200 \,\mic$, 
obtained from measurements of the Newton's law at small distances 
\cite{eotwash}, and then will be quite soon tested experimentally.

\end{document}